\documentclass[aip, 12pt, nofootinbib, superscriptaddress]{revtex4}
\usepackage{setspace}
\usepackage{amsmath, amssymb, nccmath, physics, tensor}
\usepackage{xparse, graphicx, accents}
\usepackage{mathtools, color, hyperref}
\usepackage[toc,page]{appendix}
\usepackage[mathscr]{euscript}
\usepackage{comment}

\newcommand{\ham}{\mathcal{H}}
\newcommand{\cd}{\mathcal{D}}

\newcommand{\ep}{\epsilon}
\newcommand{\kap}{\kappa}

\newcommand{\bx}{\vb{x}}

\newcommand{\by}{\vb{y}}

\newcommand{\be}{\begin{equation}}
\newcommand{\ee}{\end{equation}}
\newcommand{\ba}{\begin{eqnarray}}
\newcommand{\ea}{\end{eqnarray}}
\newcommand{\bma}{\left(\begin{array}}
\newcommand{\ema}{\end{array}\right)}

\def\bs{\begin{subequations}}
\def\es{\end{subequations}}

\def\e{\epsilon}

\def\p{\partial}

\begin{document}

\title{A Realist Interpretation of Unitarity in Quantum Gravity}
\author{Indrajit Sen}
\email{isen@clemson.edu}
\affiliation{Institute for Quantum Studies, Chapman University
One University Drive, Orange, CA, 92866, USA}

\author{Stephon Alexander}
\email{stephon\_alexander@brown.edu}
\affiliation{Brown Theoretical Physics Center, Department of Physics, Brown University, Providence, Rhode Island 02912, USA}

\author{Justin Dressel}
\email{dressel@chapman.edu}
\affiliation{Institute for Quantum Studies, Chapman University
One University Drive, Orange, CA, 92866, USA}

\begin{abstract}
Unitarity is a difficult concept to implement in canonical quantum gravity because of state non-normalizability and the problem of time. We take a realist approach based on pilot-wave theory to address this issue in the Ashtekar formulation of the Wheeler-DeWitt equation. We use the postulate of a definite configuration in the theory to define a global time for the gravitational-fermionic system recently discussed in (Phys. Rev. D 106.10 (2022): 106012), by parameterizing a variation of a Weyl-spinor that depends on the Kodama state. The total Hamiltonian constraint yields a time-dependent Schrodinger equation, without semi-classical approximations, which we use to derive a local continuity equation over the configuration space. We implement the reality conditions at the level of the guidance equation, and obtain a real spin-connection, extrinsic curvature and triad along the system trajectory. We obtain quantum corrections to deSitter spacetime from the guidance equation. The non-normalizable Kodama state is naturally factored out of the full quantum state in the conserved current density, opening the possibility for quantum-mechanical unitarity. We also give a pilot-wave generalisation of the notion of unitarity applicable to non-normalizable states, and show the existence of equilibrium density for our system. Lastly, we find unitary states in mini-superspace by finding an approximate solution to the Hamiltonian constraint.
\end{abstract}

\maketitle

\section{Introduction}

Quantum gravity effects may become important in regimes where quantum fluctuations of the gravitational field and high curvature coincide, such as close to the classical big bang and black hole singularities. In such situations and given the perturbative non-renormalizibility of quantum gravity, a non-perturbative treatment is a desired option. A conservative approach to quantization is a Schrodinger quantization, such as the Wheeler-DeWitt equation (WDW) \cite{dewitt1}. The WDW equation is non-polynomial in the metric variables and is difficult to solve. Progress was made with the Ashtekar variables which rendered the WDW equation polynomial in the configuration variables \cite{asstaker86}. \\

A major leap in progress was found by Kodama by solving the WDW equation of general relativity in terms of the Ashtekar connection \cite{kodamaginal}. This solution, called the Chern-Simons Kodama (CSK) is an exact wavefunction that solves the quantum WDW equation for a positive cosmological constant. It was shown that this Chern-Simons Kodama state consistently reduces to the Hartle-Hawking-Vilenkin state of de-Sitter space, and contains a multitude of other solutions, including black-hole quantum spacetimes \cite{Magueijo:2020ugp,genk}. Recently, an exact CSK state was found with the inclusion of fermions \cite{Alexander:2022vpn}.\\
 
 Despite this success, the CSK state as well as other formulations of the WDW equation, is fraught with conceptual and technical problems that all approaches to the WDW suffer \cite{isham93, lip, choriz}. Since time evolution is a gauge redundancy, the CSK state is timeless. Also, the Lorentzian CSK state is non-normalizable for the naive-inner product, although a recent proposal for a new non-perturbative inner product was proposed \cite{Laurent}. The twin problems of time and non-normalizability make the definition of unitarity murky. Another issue is that the non-normalizable part of the Kodama state, when linearized yields gravitons with negative energy in its spectrum. These problems are to be expected since the CSK state is background independent and a proposed ground state. In this work, we address the interconnected problems of time, normalizability and unitarity by recasting the WDW equation in the Ashtekar formalism using pilot-wave theory \cite{bohm1, bohm2, solventini, hollandbook, bell}, which is a realist formulation of quantum mechanics.\\

Our approach introduces three new ideas to attack these problems. First, we use the postulate of a definite configuration in pilot-wave to define for the first time a real, relational time in terms of variation of massless fermionic field. This allows us to discuss time evolution of the quantum state, which is shown to follow a Schrodinger equation, without using semi-classical approximations. Second, we approach the question of unitarity by deriving a continuity equation over the configuration space, instead of using operator valued reality conditions. This enables us to find a locally conserved current density on the configuration space and thereby discuss unitarity from quantum-mechanical perspective. Third, we also generalize the notion of unitarity from pilot-wave perspective, which allows us to discuss unitarity without imposing normalizability. \\

The article is structured as follows. In section \ref{sec:2} we give an introduction to the Ashtekar formalism and the Kodama state. We give an introduction to field-theoretic pilot-wave theory with a brief discussion of complex massive scalar field in section \ref{scalar}. In section \ref{main}, we develop a pilot-wave formulation of the gravitational-fermionic system in \cite{Alexander:2022vpn}, making use of some of the ideas developed in \cite{inprep}. We first introduce a global time that parameterizes a particular variation of the fermionic field and then derive the continuity equation and corresponding current density in \ref{mainc}. We discuss the physical interpretation in \ref{fizik}, including normalizability in \ref{morn}, guidance equation for the Ashtekar connection in \ref{maing}, and reality conditions in \ref{mainr}. We discuss the notion of unitarity in our approach in section \ref{prob}. We discuss quantum-mechanical unitarity using our continuity equation in \ref{qmu}, a generalized notion of unitarity using pilot-wave in \ref{pwu}, and the existence of pilot-wave unitary states in mini-superspace in \ref{probu}. We discuss our results and future directions in \ref{disc}.

\section{The Ashtekar Formalism and the CSK State \label{sec:2}}
In pursuit of a Wheeler-DeWitt quantization of gravity, it is instructive to understand how the Ashtekar connection and the resulting  Hamiltonian, diffeomorphism, and gauge constraints emerge from a manifestly covariant 4D theory of gravity.  In what follows we closely follow the derivation of the Ashtekar variables in the work of \cite{Alexander:2022vpn}. In the Ashtekar formalism \cite{asstaker86, asslake}, gravitational dynamics on a four-dimensional manifold ${\cal M}_4$ is not described by a metric $g_{\mu\nu}$ but,\footnote{We use the mostly plus metric signature, i.e. $\eta_{\mu\nu} = (-,+,+,+)$ in units of $c = 1$. We use boldface letters $\bx$ to indicate 3-vectors, and we use $x$ to denote 4-vectors. Conventions for curvature tensors, covariant and Lie derivatives are all taken from Carroll \cite{carroll04}. Greek indices ($\mu,\nu,\ldots)$ denote spacetime indices, Latin indices $(a, b,\ldots)$ denote spatial indices, and Latin indices $(I, J, \ldots)$ and $(i, j, \ldots)$ denote indices for the internal space ranging from 0, $\ldots$ 3 for the former and 1, $\ldots$ 3 for the latter.} rather, a real-valued gravitational field $e_\mu^I(x)$, mapping a vector $v^\mu$ in the tangent space of ${\cal M}_4$ at the point $x$ into Minkowski spacetime $M_4$ [with metric $\eta_{IJ} = {\rm diag}(-1,1,1,1)_{IJ}$]. Locally, the metric on ${\cal M}_4$ is $g_{\mu\nu} = \eta_{IJ}e_\mu^Ie_\nu^J$.

The Lorentz connection $\tensor{\omega}{_{\mu I}^J}$ is $\tensor{\omega}{_I^J} \equiv \tensor{\omega}{_{\mu I}^J} \dd{x^\mu}$, $\dd{\tensor{\omega}{_I^J}}\equiv \p_\mu \tensor{\omega}{_{\nu I}^J}\dd{x^\mu} \wedge \dd{x^\nu}$ is the exterior derivative, and the curvature of $\omega$ is $\tensor{R}{_I^{J}} = \dd{\tensor{\omega}{_I^{J}}} + \tensor{\omega}{_I^{K}} \wedge \tensor{\omega}{_K^{J}}$. The action of self-dual gravity is (up to the gravitational constant $8\pi G$)
\be\label{act}
S = \frac{1}{32\pi G}\int_{{\cal M}_4} \qty[*(e^I\wedge e^J)\wedge R_{IJ}  + ie^I\wedge e^J\wedge R_{IJ} - \frac{\Lambda}{6}\ep_{IJKL}e^I\wedge e^J\wedge e^K\wedge e^L]\,,
\ee
where $*$ is the Hodge dual, the first term is the Hilbert-Palatini action and the second is the Holst term (proportional to the first Bianchi identities in the absence of torsion).

Here we are interested in the Hamiltonian formulation in Ashtekar variables \cite{asstaker86, asstaker87}. In the gauge choice $e^0_\mu = 0$, it is convenient to define the densitized triad $E^{a}_{i} = \ep_{ijk}\ep^{abc} e^j_be^k_c$, which is conjugate to the self-dual connection
\be\label{sdc}
A_a^i(x) \equiv -\frac{1}{2}\tensor{\ep}{^{ij}_{k}}\tensor{\omega}{_{aj}^{k}} - i\tensor{\omega}{_{a0}^{i}}.
\ee
As the Lorentz connection (and, in particular, the spin connection $\Gamma_a^i \equiv -\tfrac12 \tensor{\ep}{^{ij}_{k}}\tensor{\omega}{_{aj}^{k}}$) is real, $A$ is complex-valued and obeys the reality conditions (for a discussion, see e.g., \cite{Kuchar93})
\be\label{rc}
A_a^i +  \overline{A_a^{i}}= 2\Gamma_a^i[E]\,,\\
E_a^i = \overline{E_a^{i}}
\ee
where $\overline{X}$ denotes complex conjugate of $X$ and the spin connection solves the equation $\dd{e} + \Gamma[E]\wedge e = 0$.

The Poisson bracket of the elementary variables $A$ and $E$ is
\be\label{comm}
\pb{A_a^i(\bx,t)}{E_j^b(\by,t)} = i8\pi G\delta_a^b\delta_j^i\delta({\bx} - {\by})\,.
\ee
Introducing the "magnetic" field and the gauge field strength
\ba\
B^{ai}&\equiv& \frac{1}{2}\e^{abc}F_{bc}^i\,,\\
F_{ab}^k &=& \p_a A_b^k - \p_b A_a^k + (8\pi G)\tensor{\e}{_{ij}^{k}}A_a^iA_b^j\,\label{F},
\ea

Now, let us construct the CSK state by solving the Wheeler-DeWitt equation.  Given the Hamiltonian
\begin{equation}
    \ham_{WDW} = \ep_{ijk}E^{ai}E^{bj}\qty(F^k_{ab} + \frac{\Lambda}{3}\ep_{abc}E^{ck}), \label{eq:ham} %= 0,
\end{equation}
which acts on some wave function $\psi[A]$, and we want to find the form of $\psi[A]$ that is annihilated by (\ref{eq:ham}). Applying the regular canonical quantization procedure, i.e.
\begin{equation}
    \hat{E}^{ai}\rightarrow 8\pi G\hbar\fdv{A_{ai}},
\end{equation}
the annihilation of the quantum state becomes
\begin{equation}
    \widehat{\ham}_{WDW}\psi[A] = (8\pi G\hbar)^2 \ep_{ijk}\fdv{A_{ai}}\fdv{A_{bj}}\qty(F^k_{ab} + \frac{8\pi G\hbar\Lambda}{3}\ep_{abc}\fdv{A_{ck}})\psi[A] = 0.
\end{equation}
\begin{comment}
If we assume that the field strength is self-dual, then $F^k_{ab} = -\frac{\Lambda}
{3}\ep_{abc}E^{ck}$ so
\end{comment} 
Putting the expression inside the brackets to zero, we get
\begin{equation}
    \ep_{abc}\fdv{\psi}{A_{ck}} = -\frac{3}{8\pi G\hbar\Lambda}F^k_{ab}\psi[A].
\end{equation}
Contracting both sides with $\ep^{dab}$ gives us
\begin{equation}
    2\delta^d_c\fdv{\psi}{A_{ck}} = -\frac{3}{\ell_{\rm Pl}^2\Lambda}\ep^{dab}F^k_{ab}\psi[A]\Leftrightarrow \fdv{\psi}{A_{ai}} = -\frac{3}{2\ell^2_{\rm Pl}\Lambda}\ep^{abc}F^i_{bc}\psi[A],
\end{equation}
where $\ell_{\rm Pl}^2 = 8\pi G\hbar$ is the Planck length. Recognizing the term multiplying the wave function to be the Chern-Simons functional, we can write down the exact solution to the Wheeler-DeWitt equation as being
\begin{equation}
    \psi_K[A] \equiv \mathcal{N}\exp(\frac{3}{2\ell_{\rm Pl}^2\Lambda}\int Y_{\rm CS}[A]),
\end{equation} 
where $\mathcal{N}$ is some normalization constant independent of the gauge field and %$Y_{\rm CS}[A]$
\begin{equation}
    Y_{\rm CS}[A] = \Tr[A\wedge\dd{A} + \frac{2}{3}A\wedge A\wedge A] = -\frac{1}{2}\qty(A^i\dd{A^i} + \frac{1}{3}\ep_{ijk}A^iA^jA^k)
\end{equation}
is the Chern-Simons functional, with the trace taken in the Lie algebra. It can be said that the WKB semiclassical limit of the CSK state is de Sitter spacetime \cite{allin},\footnote{See \cite{witten003} for criticisms of this view.} with
\be
A_a^{i} = i\sqrt{\frac{\Lambda}3}\,e^{\sqrt{\frac{\Lambda}{3}}\, t}\delta_a^i\,,\qquad E_i^a = e^{2\sqrt{\frac{\Lambda}{3}}\, t}\delta^a_i\,.
\ee

Now that we have the CSK state solely in terms of the gravitational connection and the cosmological constant, we would like to explore a full nonperturbative state that also includes the fermionic Hamiltonian.  

\section{Pilot-wave formulation of massive complex scalar field} \label{scalar}
It is helpful to begin with a discussion of pilot-wave theory with field ontology as an example (for further discussion, see \cite{bohm1, hollandbook, valentiniphd, struyvefields}). Consider a massive complex scalar field $\phi(\vec{x},t)$ on flat space-time with the Lagrangian density $\mathcal{L} = \partial^\mu\overline{\phi} \partial_\mu \phi - m^2 \overline{\phi}\phi$, where $m$ labels the mass, $\overline \phi$ labels the complex conjugate of $\phi$ and the space-time metric is $\eta = (1, -1, -1, -1)$. The conjugate momenta are $\pi = \delta \mathcal{L}/\delta \partial_0 \phi = \partial^0 \overline{\phi}$ and $\overline{\pi} = \delta \mathcal{L}/\delta \partial_0 \overline{\phi} = \partial^0 \phi$. The Hamiltonian density is $\mathcal{H} = \pi \phi + \overline{\pi}\overline{\phi} - \mathcal{L} = \overline{\pi}\pi + \vec{\nabla} \overline{\phi} \cdot \vec{\nabla} \phi + m^2 \overline{\phi}\phi$. \\

To quantize the system, the canonical commutation relations are imposed \cite{macha}. Working in the $\phi$, $\overline{\phi}$ representation, the conjugate momenta are represented by the operators $\hat{\pi} \to -i\hbar \delta/\delta \phi$, $\hat{\overline{\pi}} \to i \hbar \delta/\delta \overline{\phi}$ and the Schrodinger equation becomes
\begin{align}
    \int_\mathcal{M} \hat{\mathcal{H}} \Psi = i\hbar \frac{\partial \Psi}{\partial t}\\
    \Rightarrow \int_\mathcal{M} \qty[\hbar^2\frac{\delta^2 \Psi}{\delta \phi \delta \overline{\phi}}  + (\vec{\nabla} \overline{\phi} \cdot \vec{\nabla} \phi + m^2 \overline{\phi}\phi)\Psi ] = i\hbar \frac{\partial \Psi}{\partial t} \label{sc}
\end{align}
where $\mathcal{M}$ labels the spatial manifold and $\Psi[\phi, \overline{\phi}, t]$ is a functional of $\phi$ and $\overline{\phi}$. Using (\ref{sc}) and its complex conjugate, we can prove the following continuity equation
\begin{align}
    \frac{\partial |\Psi|^2}{\partial t} + \nabla_\phi J + \overline{\nabla}_{\phi} \overline{J}  = 0 \label{scont}
\end{align}
where $\nabla_\phi =\int_\mathcal{M} \delta/\delta \phi $ and
\begin{align}
    J = \frac{\hbar}{2i}\qty[\overline{\Psi}\frac{\delta \Psi}{\delta \overline{\phi}} - \Psi\frac{\delta \overline{\Psi}}{\delta \overline{\phi}}] = R^2 \frac{\delta S}{\delta \overline{\phi}}
\end{align}
Here $\Psi = Re^{iS/\hbar}$ and $R$, $S$ are real time-dependent functionals of $\phi$, $\overline{\phi}$. The evolution of the field is given by the guidance equation 
\begin{align}
    \frac{\delta \phi(\vec{x}) }{\delta t} \equiv \frac{J}{|\Psi|^2} = \frac{\delta S[\phi,\overline{\phi}, t]}{\delta \overline{\phi(\vec{x})}} \label{scguy}
\end{align}
Equation (\ref{scguy}) implies that the evolution of the scalar field is spatially nonlocal (over $\mathcal{M}$). This is because $S[\phi, \overline{\phi}, t]$ is a functional of $\phi(\vec{x})$, $\overline{\phi(\vec{x})}$, and thus depends on values of $\phi(\vec{x})$, $\overline{\phi(\vec{x})}$ at all $\vec{x}$ in $\mathcal{M}$ in general. Also note that (\ref{scguy}) is a local guidance equation in the configuration space $(\phi, \overline{\phi})$. That is, the evolution of a particular field $\phi(\vec{x})$ does not depend on other field configurations as $S[\phi, \overline{\phi}, t]$ in (\ref{scguy}) is evaluated at a particular point on the configuration space.

\section{Schrodinger equation of the gravitational-fermionic system} \label{main}
We wish to quantize general relativity with a positive cosmological constant and a two-component Weyl Spinor. As we will see, the corresponding total Hamiltonian constraint, which was discovered in \cite{Alexander:2022vpn}, becomes equivalent to a time dependent Schrodinger equation, where the first order spinor Hamiltonian will play exactly the role of a relational clock. This approach has advantages over scalar-field relational clocks since the latter may introduce negative norm states due to being second order in time derivative, as opposed to the Dirac equation, which is first order. \\

The Gravitational-Spinor action is:
\begin{widetext}
\begin{equation}
    S_{H+D} = \frac{1}{2\kappa}\int d^{4}x\,e(\,e^{\mu}_{I}e^{\nu}_{J}R^{IJ}_{\mu\nu} -\Lambda + i\bar{\Psi}\gamma^{I}e^{\mu}_{I}D_{\mu}\Psi -
    i\overline{D_{\mu}\Psi}\gamma^{I}e^{\mu}_{I}\Psi)
\end{equation}
\end{widetext}
where the covariant derivative is:
\begin{eqnarray}\label{covariant_derivative}
    D_{\mu}\Psi = \partial_{\mu}\Psi - \frac{1}{4}A_{\mu}^{IJ}\gamma_{I}\gamma_{J}\Psi \\
    \overline{D_{\mu}\Psi} = \partial_{\mu}\bar{\Psi} + \frac{1}{4}\bar{\Psi}\gamma_{I}\gamma_{J}A_{\mu}^{IJ}
\end{eqnarray}
upon performing a $3+1$ decomposition as discussed in section II,
the total Hamiltonian density for the combined fermionic gravitational system \cite{Alexander:2022vpn} is $\kap^{-1}(\Tilde{N} \hat{\mathcal{H}} + N^a \hat{\mathcal{V}}_a)$ where
\begin{align}
    \hat{\mathcal{H}} &= \frac{1}{2\kap}\ep_{ijk}\hat{E}^{bj}\hat{E}^{ai}\qty(F^k_{ab} + \frac{\Lambda}{3}\ep_{abc}\hat{E}^{ck}) + (\widehat{\cd}_a\xi)_A \sigma_i^{AB}\hat{E}^{ai} \widehat{\Pi}_B + \hat{E}^{ai} (\widehat{\cd}_a\xi)_A \sigma_i^{AB} \widehat{\Pi}_B \label{sup} \\
    \hat{\mathcal{V}}_a &= \frac{i}{2\kap} F^k_{ab} \hat{E}^b_k + (\widehat{\cd}_a \xi)_B \hat{\Pi}^B \label{ssup}
\end{align}
and $N > 0$, $N^a$ are the lapse function and shift vectors respectively. Here $\xi_A(\vec{x})$ is a two-component Weyl spinor and the corresponding conjugate momentum is labelled by $\Pi^B(\vec{x})$  ($A, B \in \{+, -\}$) \cite{Alexander:2022vpn}. We have chosen the Ashtekar ordering \cite{asstaker86, jacoberson, kodamaginal, allin} for the purely gravitation part of the constraints. For the interaction terms between gravity and fermion, we have Weyl ordered $\hat{E}_{ai}$ and ordered $\widehat{\Pi}_B$ to directly operate on the quantum state. We remove divergent terms throughout in our calculations.\\

The total Hamiltonian constraint is
\begin{align}
    \int_{\mathcal{M}} \textbf{ } \kap^{-1}(\Tilde{N} \hat{\mathcal{H}} + N^a \hat{\mathcal{V}}_a) \Psi[A, \xi] = 0 \label{to}
\end{align}
where $\mathcal{M}$ labels the spatial manifold. We use the quantization scheme (using commutators \cite{freedsork2, freedsork1})
\begin{equation}
    \hat{E}^{ai}\rightarrow \fdv{A_{ai}}, \hspace{0.5cm}\widehat{ \Pi}_A\rightarrow -i\frac{\delta }{\delta \xi^A}
\end{equation}
where we have used natural units. Equation (\ref{to}) implies
\begin{align}
    &\int_{\mathcal{M}} \Tilde{N} \fdv{A_{ai}}\qty[\frac{\ep_{ijk}}{2}\frac{\delta}{\delta A_{bj}}\qty(F^k_{ab} + \frac{\Lambda}{3}\ep_{abc}\frac{\delta}{\delta A_{ck}}) - 2(\widehat{\cd}_a\xi)_A \sigma_i^{AB}i \frac{\delta}{\delta \xi^B}]\Psi[A,\xi] \nonumber \\ 
    &+ \int_{\mathcal{M}} iN_b \fdv{A_{ai}}\qty[\frac{F^b_{ai}}{2} \Psi[A,\xi]]= \int_{\mathcal{M}} N^b (\widehat{\cd}_b \xi)^Bi \frac{\delta}{\delta \xi^B} \Psi[A, \xi] \label{tit}
\end{align}
Let us define
\begin{align}
    \frac{\partial \Psi[A,\xi]}{\partial t'} \equiv \int_{\mathcal{M}} \textbf{ } \frac{\delta \xi^B}{\delta t'} \frac{ \delta}{\delta \xi^B}\Psi[A,\xi] \label{partial}
\end{align}
where 
\begin{align}
    \frac{\delta \xi^B}{\delta t'} \equiv  N^b (\widehat{\cd}_b \xi)^B \label{time}
\end{align}
That is, we choose a particular variation of the fermionic field $\xi$, suggested by the form of the Hamiltonian, to implicitly define a real, formal time variable $t'$ that parameterizes this variation. Note that equation (\ref{time}) is not a semi-classical background as $\xi$ is piloted by $\Psi[A, \xi]$ via its dependence on $A_{ai}$ (see equation (\ref{guy}) below). Equation (\ref{time}) is naturally consistent with a pilot-wave interpretation as the latter posits a definite system configuration $(A, \xi)$. It is also consistent with any other interpretation where a definite configuration of the quantum system is physically meaningful.\\

We can use (\ref{time}) to rewrite equation (\ref{tit}) as
\begin{align}
     &\int_{\mathcal{M}} \Tilde{N} \fdv{A_{ai}}\qty[\frac{\ep_{ijk}}{2}\frac{\delta}{\delta A_{bj}}\qty(F^k_{ab} + \frac{ \Lambda}{3}\ep_{abc}\frac{\delta}{\delta A_{ck}}) - 2(\widehat{\cd}_a\xi)_A \sigma_i^{AB}i \frac{ \delta}{\delta \xi^B}]\Psi[A,\xi] \nonumber \\ 
    &+ \int_{\mathcal{M}} iN_b \fdv{A_{ai}}\qty[\frac{F^b_{ai}}{2} \Psi[A,\xi]] = i \frac{\partial \Psi[A,\xi]}{\partial t'} \label{schr} 
\end{align}
which resembles the time-dependent Schrodinger equation. 

\subsection{Continuity equation and current density} \label{mainc}
Using the complex conjugates of equations (\ref{partial}) and (\ref{schr}), we define (suppressing the labels $A$, $\xi$ in $\Psi$ for brevity)
\begin{align}
    \frac{\partial \overline{\Psi} \Psi}{\partial t'} = \int_{\mathcal{M}} \textbf{ }  \frac{\delta \overline{\xi^B}}{\delta t'} \frac{ \delta \overline \Psi}{\delta \overline{\xi^B}} \Psi + \int_{\mathcal{M}} \textbf{ } \overline{\Psi} \frac{\delta \xi^B}{\delta t'} \frac{ \delta \Psi}{\delta \xi^B} \label{tpartial}
\end{align}

We know from \cite{inprep} that the current density is generally of the form $|\Psi|^2 \Omega$, where $\Omega$ depends on the configuration variables and their conjugates, is independent of time, and is real and positive semi-definite. We define $\Omega \equiv \Omega[A, \overline{A}]$ as we have used $\xi$ to define our time variable $t'$. We can then show that
\begin{align}
    \frac{\partial (|\Psi|^2 \Omega)}{\partial t'} + &\int_{\mathcal{M}} \Bigg[\fdv{A_{ai}} \bigg(\Omega \overline{\Psi}\bigg\{\frac{i\Tilde{N}\ep_{ijk}}{2}\frac{\delta}{\delta A_{bj}}\bigg(F^k_{ab} + \frac{\Lambda}{3}\ep_{abc}\frac{\delta}{\delta A_{ck}}\bigg )\Psi + 2\Tilde{N} (\widehat{\cd}_a\xi)_A \sigma_i^{AB} \frac{ \delta}{\delta \xi^B} \Psi  \nonumber \\ 
    -& N_b\frac{F^b_{ai}}{2} \Psi\bigg \} \bigg) + c.c\Bigg] =  \int_{\mathcal{M}} |\Psi|^2\bigg[\frac{\delta \Omega}{\delta A_{ai}}\bigg \{\frac{i\Tilde{N}\ep_{ijk}}{2\Psi}\frac{\delta}{\delta A_{bj}}\bigg(F^k_{ab} + \frac{\Lambda}{3}\ep_{abc}\frac{\delta}{\delta A_{ck}}\bigg )\Psi \nonumber \\
    &+ \frac{2\Tilde{N}}{\Psi} (\widehat{\cd}_a\xi)_A \sigma_i^{AB} \frac{ \delta}{\delta \xi^B} \Psi + N_b\frac{F^b_{ai}}{2} \bigg \} + c.c \bigg] \label{semicontin}
\end{align}
where \textit{c.c} denotes complex conjugate of the term in square bracket, and we have used $\delta \Psi / \delta \overline{A}_{ai} = \delta \overline{\Psi}/\delta A_{ai} = 0$ $\forall  a, i$ as $\Psi$ is a holomorphic functional of $A$. The right-hand side of equation (\ref{semicontin}) can be written as
\begin{align}
    &\int_{\mathcal{M}} \qty[ \frac{\delta}{\delta A_{bj}} \bigg(\overline{\Psi} \frac{\delta \Omega}{\delta A_{ai}}\frac{i\Tilde{N}\ep_{ijk}}{2}\bigg(F^k_{ab} + \frac{\Lambda}{3}\ep_{abc}\frac{\delta}{\delta A_{ck}}\bigg )\Psi \bigg) + c.c] - \bigg[\frac{\delta}{\delta A_{ck}}\bigg(\overline{\Psi} \Psi \frac{\delta^2 \Omega}{\delta A_{ai}\delta A_{bj}}\frac{i\Tilde{N}\ep_{ijk}}{2} \frac{\Lambda}{3}\ep_{abc}\bigg) \nonumber \\
    &+ c.c\bigg] + \qty[\frac{\delta}{\delta \xi^B}\bigg ( \frac{\delta \Omega}{\delta A_{ai}} 2\overline{\Psi}\Psi\Tilde{N} (\widehat{\cd}_a\xi)_A \sigma_i^{AB} \bigg ) + c.c] + |\Psi|^2\bigg\{ -\qty[\ \frac{\delta^2 \Omega}{\delta A_{ai}\delta A_{bj}}\frac{i\Tilde{N}\ep_{ijk}}{2} F^k_{ab} + c.c] +\nonumber \\
    &\qty[\frac{\delta^3 \Omega}{\delta A_{ai}\delta A_{bj}\delta A_{ck}}\frac{i\Tilde{N}\ep_{ijk}}{2} \frac{\Lambda}{3}\ep_{abc} + c.c] - \qty[N_b\frac{\delta \Omega}{\delta A_{ai}}\frac{F^b_{ai}}{2} +c.c]\bigg\}
\end{align}

We require that $\Omega$ be such that all the source-like terms vanish. This will be true if
\begin{align}
    -\qty[\ \frac{\delta^2 \Omega}{\delta A_{ai}\delta A_{bj}}\frac{i\Tilde{N}\ep_{ijk}}{2} F^k_{ab} + c.c] +\qty[\frac{\delta^3 \Omega}{\delta A_{ai}\delta A_{bj}\delta A_{ck}}\frac{i\Tilde{N}\ep_{ijk}}{2} \frac{\Lambda}{3}\ep_{abc} + c.c]\nonumber 
    \\
    - \qty[N_b\frac{\delta \Omega}{\delta A_{ai}}\frac{F^b_{ai}}{2} +c.c] =0 \label{knoc}
\end{align}
Equation (\ref{knoc}) supplies the $\Omega$ needed to define the current density. We observe that 
\begin{align}
    \Omega[A, \overline{A}] = \frac{1}{\Psi_K[A]\overline{\Psi_K[A]}} \label{weighty}
\end{align}
solves (\ref{knoc}), where $\Psi_K[A]$ is the Kodama state. As the weight factor $\Omega$ is unique and does not depend on the Hamiltonian or the state \cite{bargman61}, we take (\ref{weighty}) henceforth. Equation (\ref{semicontin}) can then be written as
\begin{align}
    \frac{\partial (|\Psi|^2\Omega)}{\partial t'} +  \nabla^{ai} J_{ai} + \overline{\nabla}^{ai} \overline{J}_{ai} + \nabla_{B} J^{B} + \overline{\nabla}_{B} \overline{J}^{B}= 0
   \label{form}
\end{align} 
where $\nabla^{ai} \equiv \int_{\mathcal{M}} \textbf{ } \delta/\delta A_{ai}$, $\nabla_{B} \equiv \int_{\mathcal{M}} \textbf{ } \delta/\delta \xi^{B}$ and 
\begin{align}
    J_{ai} = &\frac{|\Psi|^2}{\Psi_K \overline{\Psi_K}}\bigg\{\frac{i\Tilde{N}\ell_{\rm Pl}^2}{2}\ep_{ijk}\bigg(F^k_{ab}\qty[\frac{\delta \ln \Psi}{\delta A_{bj}} + \frac{\delta \ln \Psi_K}{\delta A_{bj}}] + \frac{\ell_{\rm Pl}^2 \Lambda}{3}\ep_{abc}\bigg[\frac{1}{\Psi}\frac{\delta^2 \Psi}{\delta A_{ck}\delta A_{bj}} + \frac{\delta \ln \Psi}{\delta A_{ck}}\frac{\delta \ln \Psi_K}{\delta A_{bj}} \nonumber \\
    &- \frac{1}{\Psi_K}\frac{\delta^2 \Psi_K}{\delta A_{ck}\delta A_{bj}} + 2\frac{\delta \ln \Psi_K}{\delta A_{ck}}\frac{\delta \ln \Psi_K}{\delta A_{bj}}\bigg]\bigg ) + 2\Tilde{N}\ell_{\rm Pl}^2 (\widehat{\cd}_a\xi)_A \sigma_i^{AB} \frac{ \delta \ln \Psi}{\delta \xi^B} - N_b\frac{\ell_{\rm Pl}^2}{2\kap \hbar}F^b_{ai}\bigg \} \label{cur}\\
    J^B =& 2\frac{\ell_{\rm Pl}^2|\Psi|^2}{\Psi_K \overline{\Psi_K}}\frac{\delta \ln \Psi_K}{\delta A_{ai}}\Tilde{N} (\widehat{\cd}_a\xi)_A \sigma_i^{AB}
\end{align}
Note that equation (\ref{form}) is not yet a satisfactory continuity equation, as there are `temporal flux' terms $J^B$, $\overline{J}^{B}$ corresponding to $\xi^B$, $\overline{\xi}^B$. We can absorb them into the current density term by redefining the time parameter $t' \to t$ such that 
\begin{align}
    \frac{\delta \xi^B}{\delta t} = N^b (\widehat{\cd}_b \xi)^B +2\ell_{\rm Pl}^2\frac{\delta \ln \Psi_K}{\delta A_{ai}}\Tilde{N} (\widehat{\cd}_a\xi)_A \sigma_i^{AB} \label{ktime}
\end{align}
Equation (\ref{form}) can then be written as the continuity equation
\begin{align}
     \frac{\partial (|\Psi|^2\Omega)}{\partial t} +  \nabla^{ai} J_{ai} + \overline{\nabla}^{ai} \overline{J}_{ai} = 0 \label{qcont}
\end{align}
where $J_{ai}$ is given by (\ref{cur}). 

\subsection{Physical interpretation}\label{fizik}
Let us consider the physical interpretation given the weight factor (\ref{weighty}) and the continuity equation (\ref{qcont}). Let us first take the question of normalizability of the quantum state.
\subsubsection{Normalizability} \label{morn}
It was shown by the authors of \cite{Alexander:2022vpn}, that $\Psi[A, \xi] = \Psi_K[A]\Phi[A,\xi]$ is a good ansatz for the gravitational-fermionic WDW equation. The continuity equation (\ref{qcont}) can be rewritten as 
\begin{align}
    \frac{\partial |\Phi|^2}{\partial t} +  \nabla^{ai} J_{ai} + \overline{\nabla}^{ai} \overline{J}_{ai} = 0 \label{qqcont}
\end{align}
which makes it evident that the non-normalizable Chern-Simons-Kodama state $\Psi_K$ is factored out of the current density. Therefore, to interpret 
 (\ref{qqcont}) as a probability conservation equation, the normalizability condition is imposed on $\Phi[A,\xi]$ -- not the full quantum state $\Psi[A,\xi]$. Using (\ref{schr}), we can show that $\Phi[A,\xi]$ follows the Schrodinger equation
 \begin{align}
     &\int_{\mathcal{M}} \frac{1}{\Psi_K}\fdv{A_{ai}}\bigg \{\qty[\Tilde{N} \frac{\ep_{ijk}}{2}\frac{\delta}{\delta A_{bj}}\qty(F^k_{ab} + \frac{ \Lambda}{3}\ep_{abc}\frac{\delta}{\delta A_{ck}}) + iN_b\frac{F^b_{ai}}{2}]\Psi_K\Phi[A,\xi]\bigg\} \nonumber \\ 
    &+ \int_{\mathcal{M}}  \fdv{A_{ai}}\qty[ -2\tilde{N}(\widehat{\cd}_a\xi)_A \sigma_i^{AB}i \frac{ \delta}{\delta \xi^B} \Phi[A,\xi]] = i \frac{\partial \Phi[A,\xi]}{\partial t}
 \end{align}
 with respect to the time parameter $t$ in (\ref{ktime}). We further discuss probabilities in section (\ref{prob}).
 
\subsubsection{Guidance equations} \label{maing}
The conceptual role of the continuity equation derived from the quantum state, in pilot-wave theory, is to define the guidance equation. Using (\ref{cur}) and the standard pilot-wave prescription for the ansatz $\Psi[A, \xi] = \Psi_K[A]\Phi[A,\xi]$, the guidance equation
\begin{align}
    &\frac{\delta A_{ai}}{\delta t} \equiv \frac{J_{ai}}{|\Phi|^2} = \frac{i\Tilde{N}\ell_{\rm Pl}^2}{2}\frac{\delta \ln \Psi_K}{\delta A_{bj}} \ep_{ijk}\bigg(2 F^k_{ab} + \ell_{\rm Pl}^2 \Lambda\ep_{abc} \frac{\delta \ln \Psi_K}{\delta A_{ck}}\bigg ) - N_b\frac{\ell_{\rm Pl}^2}{2\kap \hbar}F^b_{ai} \nonumber \\
    &+ \frac{i\Tilde{N}\ell_{\rm Pl}^2}{2} \ep_{ijk}\bigg( 2F^k_{ab} \frac{\delta \ln \Phi}{\delta A_{bj}} + \frac{\ell_{\rm Pl}^2 \Lambda}{3}\ep_{abc}\qty[2\frac{\delta \ln \Phi}{\delta A_{bj}} \frac{\delta \ln \Psi_K}{\delta A_{ck}} + \frac{1}{\Phi}\frac{\delta^2 \Phi}{\delta A_{ck}\delta A_{bj}}]\bigg ) + 2\Tilde{N}\ell_{\rm Pl}^2 (\widehat{\cd}_a\xi)_A \sigma_i^{AB} \frac{ \delta \ln \Phi}{\delta \xi^B}  \label{guy}
\end{align}
determines the evolution of the Ashtekar connection with respect to the fermionic time $t$. We note that the first line of (\ref{guy}) is the classical equation of motion for the connection with $E^{bj}$ substituted by $\delta \ln \Psi_K/\delta A_{bj}$. This form of $E^{bj}$ can be shown to give the classical deSitter solution \cite{allin}. The first term in the second line of (\ref{guy}) contains the quantum corrections to the deSitter solution, whereas the second term contains the quantum contribution from the fermionic interaction with $\Pi_B$ given by $\delta \ln \Phi/\delta \xi^B$. Also note that equation (\ref{guy}) is nonlocal in the sense that the evolution of the connection at a particular point in physical space generally depends upon the value of the connection at other points in physical space, similar to equation (\ref{scguy}). \\

The guidance equation for the fermion is given by equation (\ref{ktime}). We note that (\ref{ktime}) resembles the classical equation of motion with $E^{ai}$ substituted by $\delta \ln \Psi_K/\delta A_{ai}$. However, as $A_{ai}$ is guided by the full quantum state $\Psi[A, \xi]$ in (\ref{guy}), the evolution of the fermion is implicitly state dependent and shows quantum behaviour.

\subsubsection{Reality conditions}\label{mainr}
We impose the reality conditions at the level of the guidance equation (\ref{guy}). We first note that, in the Ashtekar formulation of classical general relativity, the following conditions
\begin{align}
    A_{ai} + \overline{A}_{ai} &= 2 \Gamma_{ai} \label{cr1}\\
    E_{ai} &= \overline{E}_{ai} \label{cr2}
\end{align}
have to be imposed to recover the real sector (with real metric), where $\Gamma_{ai}$ is the 3D spin connection. In the orthodox quantum formulation of canonical quantum gravity, the reality conditions are generalised to the operator conditions
\begin{align}
    \hat{A}_{ai} + \hat{A}^{\dagger}_{ai} &= 2 \hat{\Gamma}_{ai}  \label{qr1}\\
    \hat{E}_{ai} &= \hat{E}^{\dagger}_{ai} \label{qr2}
\end{align}
We pursue here an approach based on pilot-wave theory to generalizing the classical reality conditions (\ref{cr1}), (\ref{cr2}). We demand that these conditions be met for the configuration-space trajectory determined by the guidance equation (\ref{guy}). This allows us to extract the real sector for an arbitrary solution to the Schrodinger-like equation (\ref{schr}), regardless of normalizability issues.\\

It is clear from (\ref{guy}) that the first reality condition (\ref{cr1}) will be trivially satisfied for any arbitrary solution $\Psi$ if we define
\begin{align}
2\frac{\delta \Gamma_{ai}(t)}{\delta t} \equiv \frac{\delta}{\delta t} \big (A_{ai}(t) + \overline{A}_{ai}(t)\big ) \label{spin}
\end{align}
at all points of the system trajectory. Let us next consider the second reality condition (\ref{cr2}). Using the definition 
\begin{align}
    \Gamma_{ai} = \frac{1}{2}\epsilon_{ijk}E^{bk}\qty(E^j_{a,b} - E^j_{b,a} +E^c_jE^l_aE^l_{c,b}) + \frac{1}{4}\epsilon_{ijk}E^{bk}\qty(2E^j_a \frac{\textbf{E},_b}{\textbf{E}} - E^j_b \frac{\textbf{E},_a}{\textbf{E}}) \label{solve}
\end{align}
where $\textbf{E}\equiv \det(E)$, we can solve for $E_{ai}(t)$ given $\Gamma_{ai}(t)$ along the system trajectory from (\ref{spin}). Since the $\Gamma_{ai}$ is real, (\ref{solve}) admits real solutions and the second reality condition (\ref{cr2}) is thereby satisfied. \\

Lastly, we can obtain the extrinsic curvature 
\begin{align}
    K_{ai} = \frac{1}{2\Tilde{N}}\qty( \frac{\partial N_i}{\partial x^a} + \frac{\partial N_a}{\partial x^i}  -\frac{\partial g_{ai}}{\partial t}) \label{extra} 
\end{align}
along the system trajectory from the imaginary part of the connection as
\begin{align}
    A_{ai} = \Gamma_{ai} - i K_{ai}
\end{align}

\section{Probabilities, unitarity and mini-superspace}\label{prob}

\subsection{Quantum-mechanical unitarity} \label{qmu}
Let us first consider whether the quantum-mechanical notion of unitarity is applicable. We note that since the non-normalizable $\Psi_K[A]$ is factored out of the current density in (\ref{qqcont}), it is possible that $\Phi[A, \xi] = \Psi[A, \xi]/\Psi_K[A]$ can be appropriately normalized. In that case, the continuity equation (\ref{qqcont}) may be interpreted as a statement of local probability conservation, analogous to the continuity equation in orthodox quantum mechanics. In addition, if the current $J^{ai} \to 0$ at large $|A_{ai}|$, then probabilities remain normalized\footnote{In general, the normalization of a density $\rho(\Vec{x}, t)$ evolving via the continuity equation $\frac{\partial \rho}{\partial t} + \vec{\nabla} \cdot (\rho \vec{v}) = 0$ is preserved if the current $\rho \vec{v} \to 0$ as $|x| \to \infty$.} with respect to the fermionic time and our system may be said to be quantum-mechanical unitary. We leave it to future work to determine whether such $\Phi[A, \xi]$ exist.\\

In the following, we also explore a generalised notion of unitarity that agrees with quantum-mechanical unitarity and, further, is applicable to non-normalizable $\Phi[A, \xi]$.

\subsection{Pilot-wave unitarity} \label{pwu}
The key idea here is that pilot-wave theory posits a probability continuity equation that is logically independent \cite{bohm1, bohm2, bohm54, valentinI, valentinII, teenv} of the continuity equation derived from the quantum state (\ref{qqcont}), whose role is only to define the guidance equation (\ref{guy}) for a single configuration. We may, therefore, consider an initial normalized density of configurations $\rho[A, \overline{A}, \xi, \overline{\xi}, 0]$ for a theoretical ensemble\footnote{Since pilot-wave theory has a single-world ontology, probabilities here can only refer to a single universe. For example, we can consider agents having incomplete knowledge about the universe. Such agents may assign probabilities to the possible initial configurations of the universe for a theoretical ensemble.} regardless of the normalizability of $\Phi[A, \xi]$ \cite{sen22}. The time evolution of the density is given by 
\begin{align}
    &\frac{\partial \rho[A,\overline{A},\xi, \overline{\xi}, t]}{\partial t} + \nabla^{ck}(\rho[A,\overline{A},\xi, \overline{\xi}, t] \frac{\delta A_{ck}}{\delta t})+ \overline{\nabla}^{ck}(\rho[A,\overline{A},\xi, \overline{\xi}, t] \frac{\delta \overline{A}_{ck}}{\delta t}) \nonumber\\
    & + \nabla^B(\rho[A,\overline{A},\xi, \overline{\xi}, t] \frac{\delta \xi_B}{\delta t}) + \overline{\nabla}^B(\rho[A,\overline{A},\xi, \overline{\xi}, t] \frac{\delta \overline{\xi}_B}{\delta t}) = 0 \label{physcont}
\end{align}
Equations (\ref{qqcont}) and (\ref{physcont}) imply that
\begin{align}
    \frac{d }{dt} \frac{\rho[A,\overline{A},\xi, \overline{\xi},t]}{|\Phi[A, \xi]|^2} = 0 \label{equil}
\end{align}
where 
\begin{align}
    \frac{d}{dt} &\equiv \frac{\partial }{\partial t} + \int_\mathcal{M}\frac{\delta A_{ai}}{\delta t} \frac{\delta }{\delta A_{ai}} + \int_\mathcal{M}\frac{\delta \overline{A_{ai}}}{\delta t} \frac{\delta }{\delta \overline{A_{ai}}} + \int_\mathcal{M}\frac{\delta \xi^B}{\delta t} \frac{\delta }{\delta \xi^B} + \int_\mathcal{M}\frac{\delta \overline{\xi}^B}{\delta t} \frac{\delta }{\delta \overline{\xi}^B}
\end{align}
denotes the total time derivative operator. The relation (\ref{equil}) implies that the ratio of  $\rho[A,\overline{A},\xi, \overline{\xi}, t]$ to $|\Phi[A, \xi]|^2$ remains constant along the system trajectories on configuration space. A density $\rho[A,\overline{A}, \xi, \overline{\xi}, t]$ that is equal to $|\Phi[A, \xi]|^2$ over an evolving compact support of the configuration space has been defined to be in pilot-wave equilibrium \cite{sen22}, which is a generalization of the notion of quantum equilibrium \cite{bohm54, valentinI, valentinII, teenv}. For example, an initial density (up to normalization factor)
\begin{align}
\rho[A, \overline{A},\xi, \overline{\xi}, 0] =  \label{inid}
\begin{cases}
|\Phi[A, \xi]|^2, & (A, \xi) \in \Omega_0  \\
0, & (A, \xi) \in \mathcal{C} \setminus \Omega_0
\end{cases}
\end{align}
where $\Omega_0 \equiv \{(A, \xi)| \rho[A, \overline{A},\xi, \overline{\xi}, 0] > 0\}$ is a compact support on the configuration space $\mathcal{C}$, will evolve to 
\begin{align}
\rho[A, \overline{A},\xi, \overline{\xi}, t] = 
\begin{cases}
|\Phi[A, \xi]|^2, & (A, \xi) \in \Omega_t \\
0, & (A, \xi) \in \mathcal{C} \setminus \Omega_t
\end{cases}
\end{align}
where $\Omega_t \equiv \{(A, \xi)| \rho[A, \overline{A},\xi, \overline{\xi}, t] > 0\}$ is the time evolved support on the configuration space. The behaviour of such densities has been explored for the case of harmonic oscillators in \cite{sen22}.\\

Let us next consider the notion of unitarity from a pilot-wave perspective. We define $\Phi[A, \xi]$ to be a unitary state if and only if
\begin{align}
    \lim_{|A_{ck}| \to \infty} \rho[A,\overline{A},\xi, \overline{\xi}, t] \frac{\delta A_{ck}}{\delta t} = 0 \hspace{0.2cm} \forall c,k \label{unit}
\end{align}
for any initially normalized $\rho[A, \overline{A},\xi, \overline{\xi}, 0]$ evolving via (\ref{physcont}) at any finite $t > 0$, and where $\delta A_{ck}/\delta t$ is determined from (\ref{guy}). As $\delta \xi^B/\delta t \propto \xi^B$ from (\ref{ktime}), the condition (\ref{unit}) implies that $\rho[A,\overline{A},\xi, \overline{\xi}, t]$ remains normalized with time. Clearly, the pilot-wave notion of unitarity (\ref{unit}) is applicable regardless of the normalizability of $\Phi[A, \xi]$. Note that a unitary non-normalizable state is not identical to a bound non-normalizable state \cite{sen22}. \\
 
We now explore the behaviour of solutions to the Hamiltonian constraint in the context of this discussion. This is a technically challenging question to investigate in full generality, so we address this here in the mini-superspace (FRW) approximation, which is relevant for quantum cosmology. 

\subsection{Mini-superspace}
Assuming homogenity and isotropy, we take $A_{ck}(\Vec{x}) = iA \delta_{ck}$ and $\xi^B(\Vec{x}) = \xi$. This implies that
\begin{align}
    F^k_{ab} = -\kappa A^2 \epsilon_{ab}^k \\
    (\widehat{\cd}_a\xi)_A = \kappa iA \tau_{aA}^C \xi_C
\end{align}
The Hamiltonian constraint $\hat{\mathcal{H}}\Psi = 0$ simplifies to
\begin{align}
    3i\frac{\partial^2(A^2 \Psi)}{\partial A^2} + \hbar \Lambda\frac{\partial^3 \Psi}{\partial A^3} + 2 A\tau_{aA}^C \xi_C \sigma^{aAB} \frac{\partial }{\partial A}\frac{\partial \Psi}{\partial \xi^B} + \tau_{aA}^C \xi_C \sigma^{aAB} \frac{\partial \Psi}{\partial \xi^B} = 0 \label{mex}
\end{align}
As such, equation (\ref{mex}) does not have separable solutions in $A$, $\xi$.

\subsubsection{Approximately separable solutions}
Let us make the simplifying assumption that the last term in (\ref{mex}) is small, which we will justify later. We then look for separable solutions $\Psi(A, \xi) = \chi(A) \phi(\xi)$. For such solutions, (\ref{mex}) implies
\begin{align}
    \frac{3i\phi}{A\chi'} \frac{d^2(A^2 \chi)}{d A^2} + \frac{\hbar \Lambda \phi}{A 
    \chi'} \frac{d^3 \chi}{d A^3} + 2\tau_{aA}^C \xi_C \sigma^{aAB}\frac{d \phi}{d \xi^B} = 0
\end{align}
Clearly, the first two terms depend only on $A$ whereas the third term depends only on $\xi$.
Let us introduce a separation constant $\mathcal{E}$ (in general complex) such that 
\begin{align}
    \frac{3i}{A\chi'} \frac{d^2(A^2 \chi)}{d A^2} + \frac{\hbar \Lambda}{A 
    \chi'} \frac{d^3 \chi}{d A^3} = \mathcal{E} \label{the}\\
     2\tau_{aA}^C \xi_C \sigma^{aAB}\frac{d \phi}{d \xi^B} = -\mathcal{E} \phi
\end{align}
The differential equation for $\phi$ can be written as
\begin{align}
    \xi \frac{d \phi}{d \xi} = -i\mathcal{E}_0 \phi \label{greq}
\end{align}
where $\mathcal{E}_0 = 2\mathcal{E}/ (\sigma_{aA}^+ \sigma^{aA+} + \sigma_{aA}^- \sigma^{aA+} + \sigma_{aA}^+ \sigma^{aA-} + \sigma_{aA}^-\sigma^{aA-})$ and we have used $\tau \equiv -i\sigma/2$. The general solution to (\ref{greq}) is $\phi(\xi) = c e^{-i\mathcal{E}_0\ln \xi} $, where $c$ is an arbitrary constant. We note the resemblance of this solution to the time-dependent part $e^{-iEt/\hbar}$ of an energy eigenstate corresponding to energy $E$. \\

The approximate solution to (\ref{the}) for $\chi$ is
\begin{align}
    \chi(A) = c_1 A^{-\frac{9 + \sqrt{9 + 18 i\mathcal{E} -\mathcal{E}^2} + \mathcal{E}i}{6}} + c_2 A^{-\frac{9 - \sqrt{9 + 18 i\mathcal{E} -\mathcal{E}^2} + \mathcal{E}i}{6}} \label{chi}
\end{align}
where we have neglected the third-derivative term multiplied by $\hbar \Lambda$. Note that we can select $\mathcal{E}$ in (\ref{chi}) such that the last term in (\ref{mex}) is indeed small, as assumed.

\subsubsection{Unitarity and Torsion} \label{probu}
The current (\ref{cur}) can be rewritten as
\begin{align}
    J_{ai} =|\Phi|^2\bigg\{\frac{i\Tilde{N}\ell_{\rm Pl}^2}{2}\ep_{ijk}\bigg(\frac{\ell_{\rm Pl}^2 \Lambda}{3}\ep_{abc}\frac{1}{\Psi}\frac{\delta^2 \Psi}{\delta A_{ck}\delta A_{bj}} \bigg) + 2\Tilde{N}\ell_{\rm Pl}^2 (\widehat{\cd}_a\xi)_A \sigma_i^{AB} \frac{ \delta \ln \Psi}{\delta \xi^B} - N_b\frac{\ell_{\rm Pl}^2}{2\kap \hbar}F^b_{ai}\bigg \}
\end{align}
The guidance equation (\ref{guy}) can be shown to reduce to
\begin{align}
    i \frac{dA}{dt} = -\frac{\Tilde{N}\ell_{\rm Pl}^2}{\chi}\frac{d}{dA}\bigg( \frac{\ell_{\rm Pl}^2 \Lambda}{3}\frac{d}{d A}\bigg ) \chi + 2i\hbar\Tilde{N} \mathcal{E} A \label{nimi}
\end{align}
for separable solution $\chi(A) \phi(\xi)$ corresponding to $\mathcal{E}$. Suppose that $\chi(A) = A^d$, where $d \equiv -(9 + \sqrt{9 + 18 i\mathcal{E} -\mathcal{E}^2} + \mathcal{E}i)/6$, then (\ref{nimi}) becomes
\begin{align}
     \frac{dA}{dt} = i\Tilde{N}\ell_{\rm Pl}^2 \bigg(\frac{\ell_{\rm Pl}^2 \Lambda}{3}\frac{d(d-1)}{A^2}\bigg ) + 2\hbar\Tilde{N} \mathcal{E} A \label{nimi2}
\end{align}
Clearly, for large $A$, $dA/dt$ increases approximately linearly and, using (\ref{unit}), $\chi(A) \phi(\xi)$ is pilot-wave unitary. \\

Equation (\ref{nimi2}) also implies that, in general, $A(t)$ will have both real and imaginary parts. This implies the presence of both normal and parity-violating torsion \cite{maggi1, maggi2}.

\subsubsection{Evolution of the fermionic field}
Lastly, the evolution of the fermionic field (\ref{ktime}) becomes
\begin{align}
    \frac{d\xi^B}{dt} = i\kap \qty[9\tilde{N}\kap\frac{ A^3 (\tau_{A}^{iC}) \sigma_i^{AB}}{\Lambda} + N^a  A \tau_{a}^{BC}] \xi_C \label{spinor}
\end{align}

Equation (\ref{spinor}) implies that $\xi^+$ and $\xi^-$ will quickly differ, even if $\xi^B = \xi$ at $t= 0$.

\section{Discussion} \label{disc}
We have described an interacting gravitational-fermionic system in Ashtekar formulation using the language of pilot-wave theory. We summarise here the key results of our work and potential directions for future research.\\

We have obtained a real time variable for the combined system, without semiclassical assumptions, by parameterizing variation of the fermionic field that depends on the Kodama state. In both classical and quantum canonical gravity, time disappears and the Hamiltonian becomes a constraint. Various approaches to define a natural time variable using a matter field as a clock have been discussed in the literature \cite{theo06, boho, ditti14, weasel18}. Our approach analogously uses a fermionic field to define time, but also supplements it with the pilot-wave postulate of a definite joint configuration for both the clock and the Ashtekar connection. Time is then defined to be a real variable that naturally parameterizes the variation of the definite fermionic field configuration. Both the fermionic field and the Ashtekar connection are quantized in our approach, and the time variable is well-defined for general solutions to the constraints. The total constraint is expressed as a Schrodinger equation with respect to the fermionic time. In the future, it will be interesting to explore the relationship between our approach and the previous approaches to defining time. Furthermore, our work suggests that the problem of time in quantum gravity and the problem of preferred global time required to define pilot-wave dynamics are intimately linked. Both are solved simultaneously in our approach, obviating the criticism that the preferred global time is necessarily ad hoc in pilot-wave theory. For future work, it will be interesting to apply this approach to the problem of time to scenarios with additional matter fields coupled to gravity. A straightforward application would be to vary each matter field and then sum over all to define a partial time derivative of the quantum state, in analogy with summing over the different spinor components in equation (\ref{ktime}).\\

We have derived a local continuity equation over the configuration space and discussed unitarity from both quantum-mechanical and pilot-wave perspective. It is interesting that in the context of the Tunnelling Wavefunction of the Universe, Vilenkin was able to define a conserved current for configurations in mini-superspace and it would be interesting to explore the relationship between our conserved current and his \cite{Vilenkin:1987kf,Albertini:2022yny}. In our conserved current density, the non-normalizable Kodama state is found to naturally factor out from the full quantum state. A natural question that arises for future work is whether the remaining part of the quantum state can be appropriately normalized, thereby proving quantum-mechanical unitarity. We have also given a pilot-wave generalization of the notion of unitarity, which reduces to the quantum-mechanical notion for normalizable states but is also applicable to non-normalizable states. We have shown the existence of approximate pilot-wave unitary states in mini-superspace. We leave for future work whether pilot-wave unitary states exist in general. \\

We have explored pilot-wave dynamics for the physically relevant quantities in our system. We have retrieved real spin connection, triad and extrinsic curvature along the system trajectory in configuration space by imposing the reality conditions at the level of the guidance equation for the connection. Interestingly, the guidance equation for the connection naturally resolves into the classical equation of motion, giving us deSitter spacetime as a solution, plus quantum corrections. We have also shown the existence of pilot-wave equilibrium densities \cite{sen22}, which lead to Born-rule-like probabilities. It is interesting that we have used commutators to quantize the fermionic field \cite{freedsork2, freedsork1}, and this leads to considerable simplicity in interpretation for the guidance equations. It will be interesting to explore the violation of spin-statistics theorem in quantum gravity in the future, as this is closely related to the long-standing question of particle versus field ontology for fermions in pilot-wave theory \cite{valentiniphd,struyvefields, minstrel}.\\

It is important to extract testable cosmological predictions from our approach. We know that the connection in FRW is the co-moving Horizon, $A\sim Ha$ \cite{maggi1, maggi2}, so that the evolution of the Horizon may be obtained from the guidance equation and this may yield predictions in light of the Hubble tension. Such a link would connect non-local dynamics in pilot-wave theory to the evolution of the Hubble parameter, but this is still speculative. \\

\acknowledgements
We thank Abhay Ashtekar for encouragement and technical comments on aspects of this work. We thank David Spergel for inspiring SA, years ago, to look at the de Broglie-Bohm framework in the context of cosmology. We thank Laurent Feidel, Joao Magueijo, Lee Smolin and Antonino Marciano for critical and useful comments. IS is grateful to Matt Leifer for encouragement and helpful discussions. IS was supported by a fellowship from the Grand Challenges Initiative at Chapman University.

\bibliography{bib}
\bibliographystyle{bhak}

\end{document}